\documentclass[aps,prl,twocolumn,superscriptaddress, floatfix]{revtex4}
\usepackage{amsbsy,amssymb,amsmath,bm}
\usepackage{graphicx,color,rotate}
\usepackage{units}
\usepackage{epstopdf}
\usepackage{floatrow}

\begin{document}

\title{Fermi surface of the Weyl type-II metallic candidate WP$_2$}

\author{R. Sch{\"o}nemann}\email{schoenemann@magnet.fsu.edu}
\affiliation{National High Magnetic Field Laboratory, Florida State University, Tallahassee, Florida 32306, USA}
\author{N. Aryal}
\affiliation{National High Magnetic Field Laboratory, Florida State University, Tallahassee, Florida 32306, USA}
\affiliation{Department of Physics, Florida State University, Tallahassee-FL 32306, USA}
\author{Q. Zhou}
\affiliation{National High Magnetic Field Laboratory, Florida State University, Tallahassee, Florida 32306, USA}
\author{Y. -C. Chiu}
\affiliation{National High Magnetic Field Laboratory, Florida State University, Tallahassee, Florida 32306, USA}
\author{K. -W. Chen}
\affiliation{National High Magnetic Field Laboratory, Florida State University, Tallahassee, Florida 32306, USA}
\author{T. J. Martin}
\affiliation{The University of Texas at Dallas, Department of Chemistry and Biochemistry, Richardson, TX 75080 USA}
\author{G. T. McCandless}
\affiliation{The University of Texas at Dallas, Department of Chemistry and Biochemistry, Richardson, TX 75080 USA}
\author{J. Y. Chan}
\affiliation{The University of Texas at Dallas, Department of Chemistry and Biochemistry, Richardson, TX 75080 USA}
\author{E. Manousakis}
\affiliation{National High Magnetic Field Laboratory, Florida State University, Tallahassee, Florida 32306, USA}
\affiliation{Department of Physics, Florida State University, Tallahassee-FL 32306, USA}
\author{L. Balicas}\email{balicas@magnet.fsu.edu}
\affiliation{National High Magnetic Field Laboratory, Florida State University, Tallahassee, Florida 32306, USA}

\date{\today}

\begin{abstract}

Weyl type-II fermions are massless quasiparticles that obey the Weyl equation and which are predicted to occur at the boundary between electron- and hole-pockets in certain semi-metals, i.e. the (W,Mo)(Te,P)$_2$ compounds. Here, we present a study of the Fermi-surface of WP$_2$ \emph{via} the Shubnikov-de Haas (SdH) effect. Compared to other semi-metals WP$_2$ exhibits a very low residual resistivity, i.e. $\rho_0 \simeq 10$ n$\Omega$cm, which leads to perhaps the largest non-saturating magneto-resistivity $(\rho(H))$ reported for any compound. For the samples displaying the smallest $\rho_0$, $\rho(H)$ is observed to increase by a factor of $2.5 \times 10^{7}$ $\%$ under $\mu_{0}H = 35$ T at $T = 0.35$ K. The angular dependence of the SdH frequencies is found to be in very good agreement with the first-principle calculations when the electron- and hole-bands are slightly shifted with respect to the Fermi level, thus supporting the existence of underlying Weyl type-II points in WP$_2$.

\end{abstract}

\maketitle


Weyl fermions are predicted to emerge as low energy excitations in semimetals characterized by strong spin-orbit coupling and lack of inversion or time-reversal symmetries \cite{soluyanov_type-ii_2015,felser,bernevig2,Weyl1,Weyl2, Hasan,xu_discovery_2015}. Two types of Weyl semi-metallic systems (WSM) have been proposed: Type-I displays linearly dispersing bands which cross at pairs of point-like Fermi surfaces (or Weyl points), while in type-II the Weyl points appear at linearly touching points between electron and hole pockets \cite{soluyanov_type-ii_2015} resulting from the intersection of these bands with the Fermi level. Weyl points act as a topological charges, or as either sources or sinks of Berry-phase curvature pseudospin. Experimental fingerprints of WSM systems are topological Fermi arcs on the Fermi surface (FS) of the surface states \cite{xu_discovery_2015, deng_experimental_2016, tamai_fermi_2016} and unconventional magnetotransport properties due to the Adler-Abel-Jackiw anomaly \cite{adler,bell,nielsen},
which corresponds to the pumping of charge carriers between Weyl points of opposite charge, or chirality, under the simultaneous presence of electric and magnetic fields
\cite{adler,bell,nielsen,zyuzin_topological_2012, burkov_chiral_2015, huang_observation_2015, zhang_signatures_2016}.

\begin{figure*}
\begin{center}
		\includegraphics[width = 14 cm]{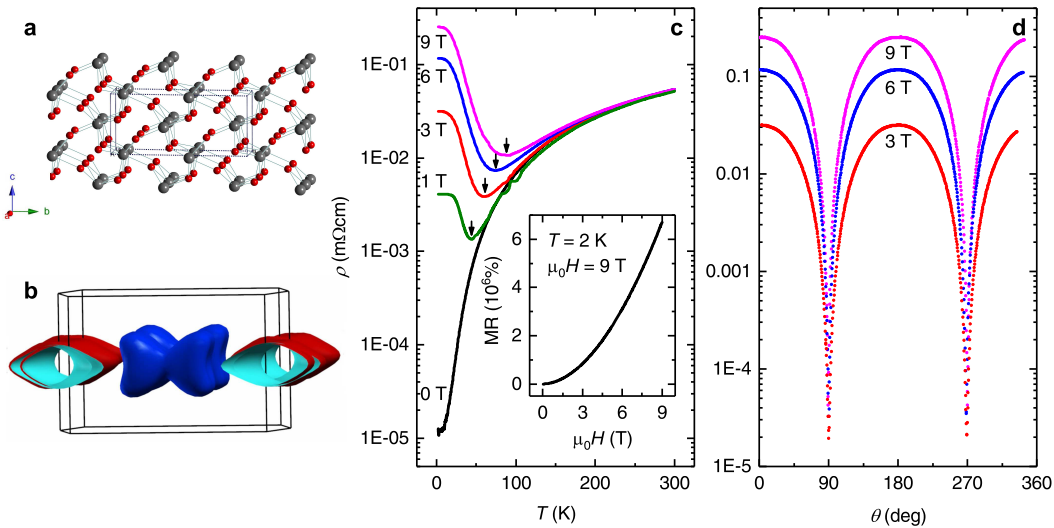}
		\caption{(a) Crystallographic structure and orthorhombic unit cell of $\beta$-WP$_2$. The W and P atoms are depicted in grey and in red, respectively. (b) Fermi surface of WP$_2$, which consists of two pairs of electron- (blue) and hole-pockets (red) within the first Brillouin zone. (c) Resistivity $\rho$, from a WP$_2$ single-crystal having a RRR of 4750, as a function of the temperature $T$ under several field values. The arrows indicate the ``turn-on" temperature given by the minimum in $\rho(\mu_0H, T)$. Inset: for this particular sample $\rho(\mu_0H)$ exceeds $6 \times 10^{6}$ \% under $T = 2$ K and $\mu_{0}H = 9$ T. (d) Angular-dependence of the magnetoresistivity at $ T = 2\,\mathrm{K}$. The field is rotated in a plane perpendicular to the crystallographic $a$-axis where $\theta = 0^{\circ}$ corresponds to $H\parallel b-$axis.}
	\label{fig:panel1}
\end{center}
\end{figure*}

Te based transition-metal dichalcogenides such as WTe$_2$ and orthorhombic MoTe$_2$ (or $\gamma-$MoTe$_2$) were predicted to display a Weyl type-II semi-metallic state \cite{felser, bernevig2}. A series of angle-resolved photoemission spectroscopy (ARPES) experiments \cite{tamai_fermi_2016, arpes} on  MoTe$_2$ claim to observe topological Fermi arcs and a FS whose geometry is in broad agreement with first-principle calculations \cite{bernevig2}. However, this contrasts with a quantum oscillatory study which reveals a far simpler Fermi surface displaying three-dimensional character \cite{daniel} despite the layered nature of this compound. The reason for such a discrepancy remains unclear, but it does question the validity of both the ARPES measurements and the predictions by density-functional theory calculations. Recently, a type-II WSM ground-state was predicted for the transition-metal diphosphides MoP$_2$ and WP$_2$ \citep{autes_robust_2016}. WP$_2$ crystallizes in two distinct structures \citep{ruhl_uber_1983, martin_synthesis_1990, mathis_reduction_1991}: the $\alpha$-phase which displays the OsGe$_2$ structure type and the $\beta$-phase which is isostructural to MoP$_2$ and belongs to the non-centrosymmetric orthorhombic space group $Cmc2_{1}$ (36) \citep{ruhl_uber_1983} (see Fig. \ref{fig:panel1}(a)) and is  predicted to host Weyl points. Unlike WTe$_2$ and $\gamma-$MoTe$_2$, $\beta-$WP$_2$ is not a layered material which leads to a simpler band structure and a more robust arrangement of the Weyl points with respect to small structural changes \citep{autes_robust_2016}. Based on DFT calculations it was shown that the electron and hole pockets touch at two inequivalent points located at $0.471$ and $0.340$ eV below the Fermi energy $E_{\mathrm{F}}$ resulting in a total of eight Weyl points across the $k_{x}$-$k_{y}$-plane \citep{autes_robust_2016}. In addition, a very recent report highlights its unusual electrical transport properties including extremely large magnetoresistivity (MR) possibly related to its underlying Weyl physics \cite{kumar_extremely_nodate}.

In this letter we present a detailed study on the Fermi surface of $\beta$-WP$_2$ \emph{via} measurements of the Shubnikov-de Haas (SdH) effect which allows us to compare the geometry of the FS determined experimentally with the one predicted by the DFT calculations. The SdH-effect can only detect the electronic structure at $E_{\mathrm{F}}$ and is not able to directly probe the existence of the Weyl points below $E_{\mathrm{F}}$. But a good agreement between the calculated and the experimentally determined Fermi surfaces validates the band-structure calculations and therefore the predicted existence of underlying Weyl type-II points in this compound. Given that previous quantum oscillatory measurements \cite{daniel} unveiled strong discrepancies with the DFT predictions, our findings make $\beta$-WP$_2$ one of the most promising candidates for the realization of Weyl quasiparticles.

WP$_2$ single crystals were grown through a chemical vapor transport (CVT) method as described in detail within the Supplemental Information (SI) file \cite{Supplemental}.
Conventional four-terminal resistivity measurements were performed in a physical property measurement system under magnetic fields up to $\mu_0H = 9$ T and temperatures as low as $T = 2$ K. The angular dependence of the Shubnikov-de Haas effect under magnetic fields up to $35\,\mathrm{T}$ was measured in a resistive Bitter magnet at the National High Magnetic Field Laboratory in Tallahassee. For further details see SI. Band structure calculations were performed by using the Quantum Espresso \cite{giannozzi_quantum_2009} implementation of the Density Functional Theory (DFT) in framework of the Generalized Gradient Approximation (GGA) including spin-orbit coupling (SOC). The Perdew-Burke-Ernzerhof (PBE) exchange correlation functional \citep{hamann_optimized_2013} was used with fully relativistic, norm conserving pseudo-potentials generated by using the optimized norm-conserving Vanderbilt pseudo-potential as described in Ref. \citep{hamann_optimized_2013}. For additional information see SI. The lattice parameters used in the calculations were extracted from single-crystal X-ray diffraction measurements and are in good agreement with previous reports \cite{noauthor_materialsproject_nodate}, see SI.

\begin{figure}[htb]
\begin{center}
		\includegraphics[width = 7.5 cm]{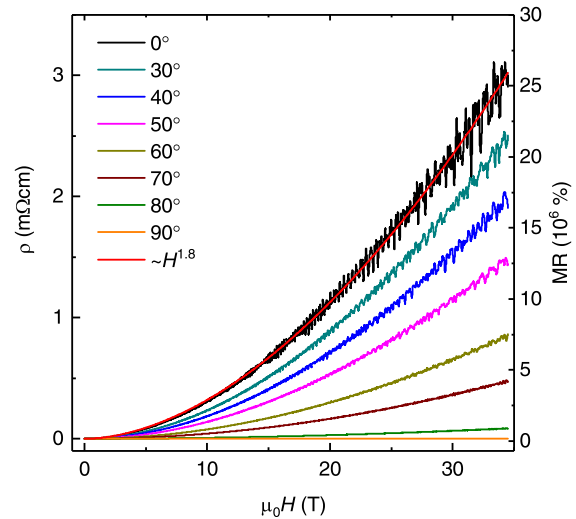}
		\caption{$\rho$ as a function of $\mu_0 H$ for a WP$_2$ single-crystal at $T = 0.35\,\mathrm{K}$ and for several angles between $\theta = 0^{\circ}$ (or $H\parallel b-$axis) and $90^{\circ}$ ($H\parallel a$). The amplitude of the oscillatory component superimposed onto $\rho(H)$, or the SdH-effect, is most pronounced for $\mu_0H \parallel b$-axis, whereas no oscillation can be resolved for $\mu_0H \parallel a$. $\rho(\mu_0H)$ can be described by a single power law $\rho(H)\propto H^{\lambda}$ for $\mu_{0} H > 0.5\,\mathrm{T}$, with $\lambda=1.8$ for $H\parallel b$ (red line) and $\lambda \approx (1.85 \pm 0.05)$ for $\theta\neq 0^{\circ}$.}
	\label{fig:panel2}
\end{center}
\end{figure}
\begin{figure*}[htp]
	\centering
		\includegraphics[width = 14 cm]{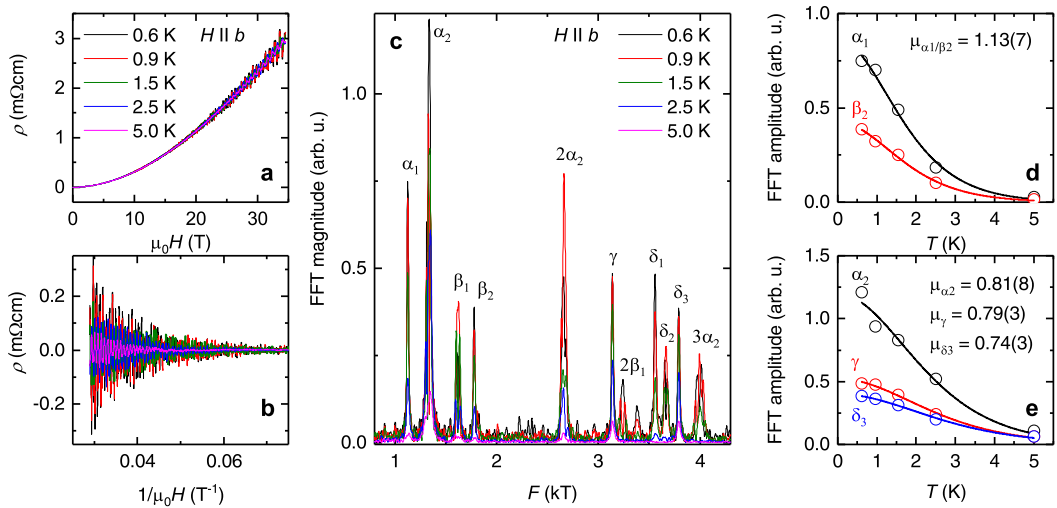}
		\caption{(a) $\rho$ as a function of $\mu_0H \parallel b-$axis for several temperatures. (b) Oscillatory component superimposed onto $\rho$, after subtraction of the background magnetoresistivity, as a function of $(\mu_0H)^{-1}$. (c) Fast Fourier Transform (FFT) of the SdH oscillations/signal shown in (b). The observed peaks and their second harmonics are labeled upon identification of the associated Fermi surface sheets. $\alpha$ and $\beta$ peaks result from cyclotronic orbits on the hole-pockets, while $\gamma$ and $\delta$ results from orbits on the electron sheets. Electron and hole-sheets are split by the spin-orbit coupling. (d, e) Amplitude of the FFT peaks as a function of $T$ where solid lines represent fits to the Lifshitz-Kosevich formalism from which one extracts the effective cyclotron masses.}
	\label{fig:panel3}
\end{figure*}
Figures \ref{fig:panel1}(a) and \ref{fig:panel1}(b) display respectively, the crystallographic structure of WP$_2$ and its calculated FS within the first BZ.
The electrical resistivity $\rho$ as a function of the temperature $T$ under magnetic fields $\mu_0H=0, 1, 3, 6,$ and 9 T are shown in Fig. \ref{fig:panel1}(c). For all of the measured samples the current $j$ was injected along the $a$-axis while the field remained perpendicular to $j$ by rotating it between the $c-$ and the $b-$axes. The crystal whose data is shown in the Fig. \ref{fig:panel1}(c) displays a residual resistivity ratio (RRR) of 4750 with a residual resistivity $\rho_{0}$ of $\approx 10$ n$\Omega$cm. This small value is unique among transition-metal chalcogenides/pnictides but comparable to those observed in pure metals \cite{noauthor_materialsproject_nodate}. For $\mu_{0}H = 9\,\mathrm{T}$ applied along the $b$-axis the magneto-resistivity (MR) exceeds $6 \times 10^{6}$ \% at $T = 2\,\mathrm{K}$, which is comparable to WTe$_2$ under $\mu_0H = 60$ T \citep{ali_large_2014}. For all of the studied crystals, the RRR was found to vary between 1000 and $\approx 20000$. For subsequent measurements we selected crystals with  $4000 \leq $ RRR $ \leq 10000$, finding no variations in the SdH signal superimposed onto $\rho (\mu_0H)$ among the different samples.

The FS shown in Fig. \ref{fig:panel1}(b) consists of 2 pairs of electron- and hole-pockets which, in absence of inversion symmetry, are split by the spin-orbit coupling. The electron-pockets are closed while the hole ones are corrugated cylinders responsible for the marked anisotropy in $\rho(\theta, \mu_0H)$ shown in Fig. \ref{fig:panel1}(d), which results from cyclotronic and open orbits on the FS. $\theta$ is the angle between $\mu_0H$ and the $b-$axis. The anisotropy in $\rho(\theta, \mu_0H)$ observed for fields along the $b$-axis and fields along the other two axes reaches $\sim 6000$ under $T = 2\,\mathrm{K}$ and $\mu_{0}H = 9\,\mathrm{T}$ and $\sim 35000$ under $T = 0.35\,\mathrm{K}$ and $\mu_{0}H = 35\,\mathrm{T}$. $\rho(\mu_0 H)$ follows an anomalous power-law when $\mu_0H \gtrsim 0.5\,\mathrm{T}$, i.e. $\rho(\mu_0 H)\propto (\mu_0 H)^{\lambda}$, with $\lambda = 1.8$, for $\mu_0 H\parallel b$ and $\lambda \approx 1.8 - 1.9$ for the other field orientations. No saturation was observed in $\rho(\theta, \mu_0H)$ for fields up to $35\,\mathrm{T}$, in agreement with an earlier report indicating no saturation up $60\,\mathrm{T}$ \cite{kumar_extremely_nodate}. Hall effect measurements, analyzed via a two-band model and discussed in the SI \cite{Supplemental}, confirm that WP$_2$ is a well-compensated conductor, which explains its pronounced magnetoresistivity. Subsequently, we focus on the quantum oscillatory phenomena or on the pronounced SdH-effect superimposed onto the raw data plotted in Fig. \ref{fig:panel2}. As the field is rotated from the $b-$ towards the $a$-axis the magnitude of the SdH oscillations decreases continuously and become unobservable for fields along the $a$-axis. However, we were able to observe a decrease in the longitudinal magnetoresistivity within a narrow angular window around the $a-$axis (see, SI), when the current is aligned along the magnetic field. Whether this results from the chiral anomaly \cite{adler,bell,nielsen,zyuzin_topological_2012, burkov_chiral_2015, huang_observation_2015, zhang_signatures_2016} or from current jetting (see Refs. \cite{pippard_magnetoresistance_2009, ueda_anisotropy_1980}) remains unclear and will require additional studies. In WP$_2$, the effects of the chiral anomaly associated with its Weyl type-II points are predicted to be observable when fields and currents are parallel to the $b$-axis which, as already pointed out by Ref. \cite{kumar_extremely_nodate}, is an experimentally challenging task.

\begin{figure*}[htp]
	\centering
		\includegraphics[width = 15 cm]{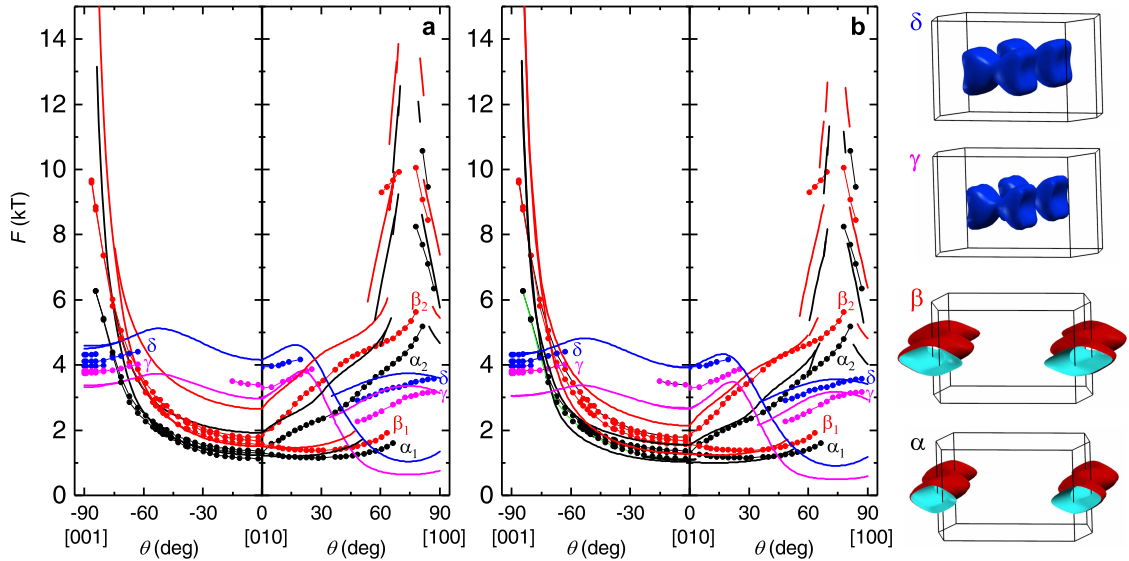}
		\caption{SdH frequencies $F$ observed in WP$_2$ as a function of the angle $\theta$. Markers indicate the position of the peaks observed in the FFT spectra while solid lines depict the angular dependence of the frequencies predicted by the DFT calculations. In (a) the predicted frequencies are shown without a displacement of the Fermi level. (b) Predicted $F(\theta)$ after the Fermi level has been shifted to improve the agreement with the experiments, see main text.  Colors of markers and lines indicate the associated Fermi surface pockets (shown to the right). Black and red ($\alpha$ and $\beta$) represent the frequencies associated with the spin-orbit split electron pockets. Magenta and blue ($\gamma$ and $\delta$) represent orbits on the hole pockets.}
	\label{fig:panel4}
\end{figure*}
Figure \ref{fig:panel3} (a) shows $\rho$ as a function of $\mu_0H \parallel b$-axis for several temperatures $T$ ranging from 0.6 to 5.0 K. Figure \ref{fig:panel3}(b) displays the superimposed oscillatory signal after subtraction of the magnetoresistive background. The Onsager relation associates every frequency $F$ to a FS extremal cross-sectional area $A$.
The fast Fourier transform (FFT), or the frequency spectra of the SdH signal, are shown in Fig. \ref{fig:panel3}(c).  The observed FFT peaks can be assigned to individual FS pockets by comparing their positions with the frequencies predicted by our DFT calculations: the $\alpha$ and $\beta$ peaks belong to the hole-like sheets, the $\gamma$ and the $\delta$ ones to the electron pockets. Usually, there is more than just one extremal cross-sectional orbit (i.e. maxima and minima), hence multiple peaks can be assigned to a single FS sheet. Therefore, we proceed by labeling these orbits as the $\alpha_{1, 2}$ (or the $\beta_{1,2}$, etc.) frequencies. The effective cyclotron masses $\mu$ of the associated charge carriers can be extracted from the amplitude of the peaks seen in the FFT spectra as a function of the temperature. Their $T-$dependence is described by the Lifshitz-Kosevich temperature damping factor \cite{shoenberg_magnetic_1984}, i.e. $R_{T} = BT/\sinh(BT)$, where $B = 14.69/\overline{\mu_0H}\mu$ with $\overline{\mu_0H}$ being the average inverse field. For the hole pockets we obtain $\mu$s between $\approx 0.8$ and 1.1 $m_0$, where $m_0$ is the bare electron mass, and $0.7-0.8$ $m_0$ for the electron pockets. For $\mu_0H \parallel a$-axis we obtain considerable heavier masses, between $3$ and $10$ $m_0$ for the electron pockets. See, SI for a comparison between  calculated and  experimentally extracted frequencies and effective masses.

Figures \ref{fig:panel4}(a, b) display the angular-dependence of the SdH frequencies obtained by rotating $\mu_0H$ in the $a-b$ and in the $b-c$ planes.  The position of the peaks in the FFT spectra are shown as markers, whereas the theoretically predicted frequencies are indicated by solid lines. Higher harmonics of the fundamental frequencies were omitted for clarity. In Fig. \ref{fig:panel4}(b), the calculated electron- and hole-pockets were displaced, respectively, by $+30\,\mathrm{meV}$ and $-30\,\mathrm{meV}$ relative to $E_{F}$. This improves significantly the agreement between the calculated and the measured frequencies, particularly in what concerns the $\alpha_{2}$ and the $\beta_{2}$ branches. The two-dimensional character of the hole-pockets ($\alpha$, $\beta$ branches) is indicated by the divergence of the SdH frequencies upon approaching the $a$- or the $c$-axis. In contrast, the frequencies related to the electron pockets remain finite within the whole angular range. Large portions of the FFT spectra are dominated by the hole-frequencies, making it difficult to track the position of the much less pronounced FFT peaks associated with the electron pockets (i.e. $\gamma$, $\delta$), since they seem to partially overlap higher harmonics of the $\alpha$ and $\beta$ frequencies. SdH frequencies can be assigned to the $\gamma$ and $\delta$ branches only for angles close to the $a-$, $b-$ and the $c-$ axes. This is the reason for the absence of markers between $\theta = -60^{\circ}$ and $-20^{\circ}$ for rotations in the $b-c$ plane. Furthermore, for $F \lesssim 2$ T we were not able to detect any of the electron-like orbits predicted by DFT for rotations in the $b-c$ plane. Nevertheless, the FFT peaks assigned to the electron pockets display a good agreement with the theoretically predicted ones (depicted by blue and magenta lines).

In summary, WP$_2$ displays a very low residual resistivity and a gigantic magneto-resistivity for magnetic fields applied along its $b-$axis. These are clear indications for a very clean compound which, according to our Hall-effect measurements, is carrier compensated. The magnetoresistivity decreases by several orders of magnitude when the field is rotated towards the $b-$axis. This suppression results from two factors: open orbits on the hole-Fermi surfaces, and remarkably heavy effective masses (or lower mobilities) for the carriers performing cyclotronic motion on the electron-pockets. More importantly, our detailed study on the Shubnikov-de Haas effect reveals a Fermi surface geometry in quite good agreement with first-principle calculations. Density Functional Theory finds that WP$_2$ is a Weyl type-II ``semi-metal", hence our results support the calculations. Although the Weyl type-II points are located well-below the Fermi-level, the observation of negative longitudinal magneto-resistivity might indicate that these can influence carriers at the Fermi level \emph{via} the axial anomaly. The large number of SdH frequencies observed here precludes a reliable extraction of the Berry phase by fitting the oscillatory signal to the Lifshitz-Kosevich formalism. Hence, to confirm the topological character of this compound, angle resolved photoemission experiments would have to be performed in order to detect  Fermi arcs \cite{soluyanov_type-ii_2015,felser,bernevig2,Weyl1,Weyl2, Hasan,xu_discovery_2015}. Finally, we want to point out that the density-of-carriers in WP$_2$ is 2 orders of magnitude larger those of conventional semi-metals, in agreement with its rather large Fermi surfaces. Hence, this compound should be classified as being metallic instead of ``semi-metallic".

This work was supported by DOE-BES through award DE-SC0002613. The NHMFL is supported by NSF through NSF-DMR-1157490 and the State of Florida. Correspondence and requests for materials should be addressed to R.S. or to L.B.





\end{document}